\documentclass[%
 reprint,
 amsmath,amssymb,longbibliography,aps,prb]{revtex4-2}
\pdfoutput=1
\usepackage{graphicx,multirow}
\usepackage{dcolumn}
\usepackage{bm}
\usepackage{pdfpages} 
\usepackage{pgffor} 

\newcolumntype{k}{>{\columncolor{blue!20}}c}
\newcolumntype{r}{>{\columncolor{blue!10}}c}
\newcolumntype{d}{>{\columncolor{red!10}}c}

\newcommand{\spin}{\sigma}
\newcommand{\ospin}{\overline{\sigma}}

\newcommand{\kv}{{\bf k}}
\newcommand{\qv}{{\bf q}}

\newcommand{\q}{{\bf q}}

\makeatletter
\AtBeginDocument{\let\LS@rot\@undefined}
\makeatother

\def\supplementfilename{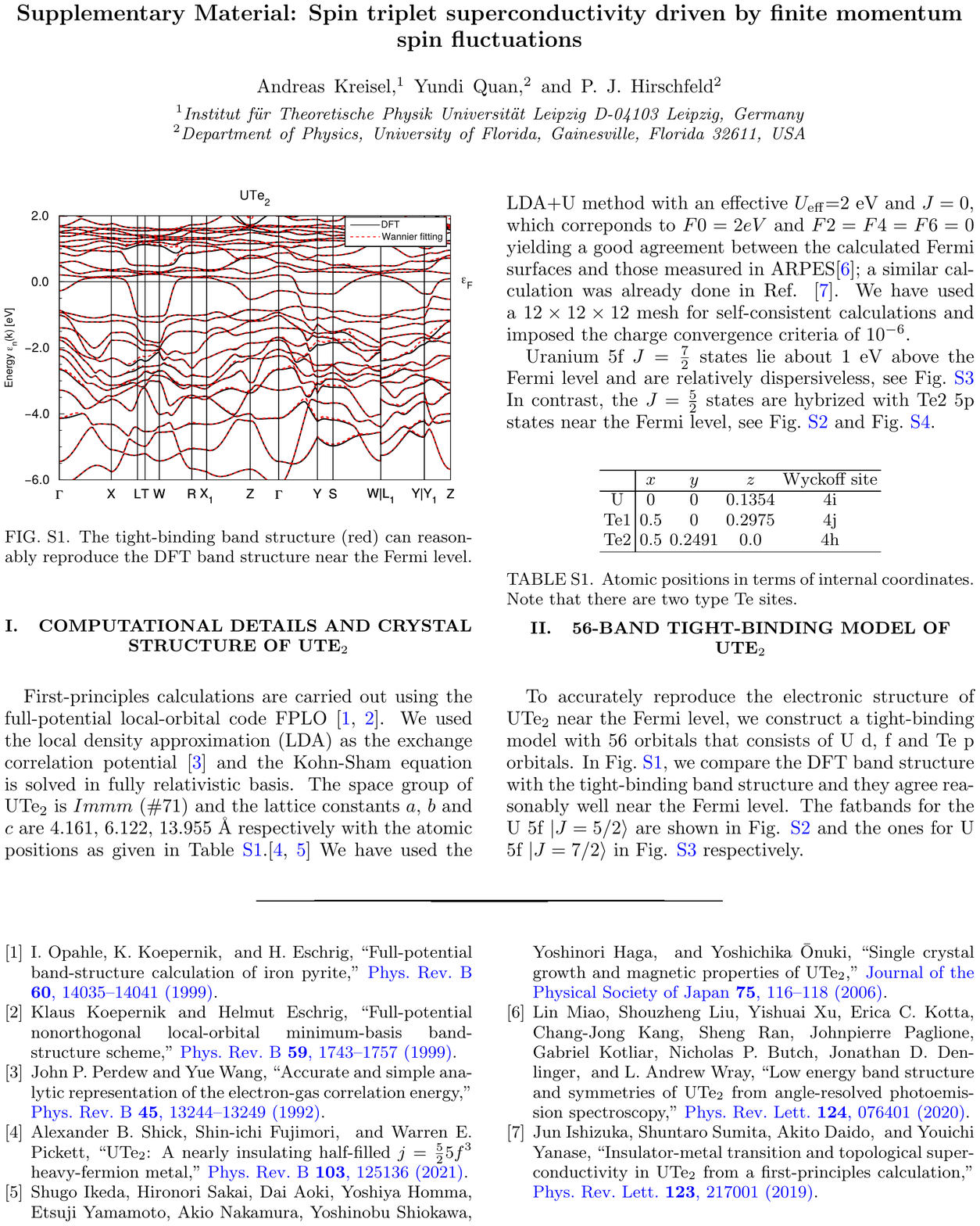}

\pdfximage{\supplementfilename}
\def\numbersupplementpages{\the\pdflastximagepages}

\newif\ifarXiv
\arXivtrue 

\usepackage{hyperref}

\begin{document}

\title{Spin triplet superconductivity driven by finite momentum spin fluctuations}

\author{Andreas Kreisel$^1$, Yundi Quan$^2$,  P. J. Hirschfeld$^2$}
\affiliation{
$^1$Institut f\"ur Theoretische Physik Universit\"at Leipzig
	D-04103 Leipzig, Germany\\
$^2$Department of Physics, University of Florida, Gainesville, Florida 32611, USA
}%


\begin{abstract}
A small number of superconductors are believed to exhibit intrinsic spin triplet pairing, and they are often discussed in terms of a simple, $^3$He-like picture where ferromagnetic spin fluctuations provide the ``glue".  However, in some cases in which reliable inelastic neutron scattering measurements are available, spin excitations are found to be peaked at finite momentum $\qv$ rather than $\qv=0$. Here we investigate some simple models that exhibit triplet pairing arising from antiferromagnetic spin fluctuations. We show that a strong peak at larger $\qv$ in the magnetic susceptibility can drive such states and can give rise to pairing states with nodes in the $k_z$ plane even in the presence of a pure 2D Fermi surface. In  these situations,  dominant pair scattering processes occur between Fermi surface segments with like signs of the superconducting order parameter, yet they are consistent with an overall odd parity state.   We examine the applicability of these scenarios to putative triplet superconductors UTe$_2$ by calculations based on three dimensional Fermi surfaces.
\end{abstract}

\maketitle


\section{\label{sec:intro} Introduction}

The importance of the effect of spin fluctuations on superconductivity has been recognized since the early work of Berk and Schrieffer\cite{berk} and Doniach and Engelsberg\cite{Doniach1966},
who pointed out that unexpectedly low measured critical temperatures in the transition metal series could be understood by accounting for these effects, which contribute
a repulsive term to the effective pairing interaction.  It was recognized by  Layzer and Fay\cite{LayzerFay1971} that the same  interaction would be attractive in the triplet channel.  These calculations were performed for an electron gas, where the magnetic susceptibility $\chi_0(\q)$ was peaked at wavevector $\q=0$, and strongly enhanced near the Stoner instability $U\chi_0(\q)=1$, where $U$ is a local repulsive interaction. Later, it was realized that unconventional singlet pairing can be favored if the same susceptibility is peaked at nonzero wavevector, e.g. as obtained for tight-binding models near half-filling\cite{Cyrot1986,Scalapino1986,Miyake86}.  For example, finite-momentum spin fluctuations in cuprates near $(\pi,\pi)$ are often thought to drive a $d$-wave state, which changes sign over the Fermi surface in order to take advantage of the repulsive interaction\cite{Scalapino1995}. 

Based on the preceding considerations, it is generally considered natural for ferromagnetic spin fluctuations to drive triplet (odd-parity) pairing, while antiferromagnetic fluctuations should drive singlet (even-parity) pairing.  Recently, an apparently ``unnatural" situation has arisen in the novel unconventional superconductor UTe$_2$.  This  heavy fermion compound is now believed to be a spin triplet superconductor on the basis of NMR experiments and a thermodynamic phase diagram including reentrant phases at high fields quite similar to known ferromagnetic superconductors\cite{Ran2019,Aoki2019}.  However, inelastic neutron scattering experiments  found spin excitations primarily at intermediate incommensurate wavevectors\cite{Dai2020}, and under applied pressure, an antiferromagnetic, rather than ferromagnetic phase is nucleated at 1.3GPa\cite{Rosa2020}.   At present the physical pairing mechanism in this material is the subject of considerable debate, and the pairing channel itself is not yet determined, so it may well be that spin fluctuations are not responsible for superconductivity in the usual sense.  It is, however, also possible that in systems like UTe$_2$ unusual aspects of the electronic structure, in particular its lower symmetry, may conspire to produce a pairing interaction that is odd parity with a susceptibility peaked at nonzero $\q$.

There may already be an example of the latter phenomenon.  While the  order parameter symmetry in the canonical heavy fermion superconductor UPt$_3$ is not completely agreed upon, there is considerable evidence in favor of a triplet $E_{2u}$ representation\cite{Sauls2018}; yet inelastic neutron experiments find peaks at nonzero wave vector\cite{Aeppli1987,Hayden1992,vanDijk2002}.    A three-dimensional (3D) microscopic spin fluctuation solution of the linearized gap equation indeed found such a state to be stable, but it is not clear if it is competitive with singlet states. Notably, the actual gap structure identified was considerably more complicated than those considered in earlier, phenomenological calculations\cite{Ikeda2016}.  The band structure used indeed led to a susceptibility peaked at wave vectors similar to those observed in the  neutron measurements.  

It is thus quite interesting to explore the general question of when a spin triplet paired state can be favored in the presence of strong nonzero momentum  scattering in order to understand the pairing mechanism in these two materials and guide the search for other intrinsic triplet superconductors.  In this paper we begin with the simplest possible models with repulsive interactions, and we ask how the electrons pair given a magnetic susceptibility function with assumed peaks at $\q\ne 0$, consistent with the symmetry of the lattice.  We begin with simple 2D models, and we show that nonzero-$q$ triplet pairing is already possible. Then we discuss models with dispersion in the third direction, and how they can be optimized.  Finally, we discuss the case of UTe$_2$  in the context of what we have learned of the conditions required for finite-$q$ triplet pairing.

To our knowledge, the general problem of how to generate triplet pairing with finite $q$ spin fluctuations has only been discussed before in the context of systems with significant spin-orbit coupling spin anisotropy.  For example, several authors attempted to explain chiral $p$-wave superconductivity in  Sr$_2$RuO$_4$\cite{Ogata2000,Sato2000,Kuroki2001a} using this approach, assuming magnetic susceptibilities $\chi_{ij}(\q)$, where $i$ and $j$ are directions in spin space, resembling results from density functional theory (DFT) calculations, but additionally allowing for a spin anisotropy such that $\chi_{+-}/\chi_{zz}<1$ and the overall magnitude of the peaked structure of both components at a Fermi surface nesting vector. Other authors investigated similar scenarios in the context of organic superconductors\cite{Kuroki2001b}.

In multiband systems, however, the susceptibility peak is not necessarily peaked at wave vectors determined by the Fermi surface.  We therefore adopt a more general phenomenology, and show that finite-$q$ triplet 
pairing is possible even in the absence of the spin-orbit coupling necessary to enhance the anisotropy in the spin response.

\section{Model}

We consider a single band tight binding model with the Hamiltonian
\begin{equation}
 H_0=\sum_{\kv \spin}\xi_{\kv} c_{\kv\spin}^\dagger c_{\kv\spin},
 \label{eq_tb}
\end{equation}
where $\xi_\kv$ is the Fourier transform of the hopping matrix elements which for a square lattice system in two dimensions would be given by $\xi_\kv=-2t[\cos(k_x)+\cos(k_y)]-4t'\cos(k_x)\cos(k_y)-\mu$ with $t$ being the hopping integral to nearest neighbors, and $t'$ is the hopping integral between next-nearest neighbors.

When such a system is subjected to interactions, it can exhibit an instability to a superconducting state such that the superconducting gap is given by the self consistent solution of the equation\cite{Romer2015}
\begin{eqnarray}
  \Delta^{s/t}_\kv&=&-
  \frac{1}{2N}\sum_{\kv'}\Gamma^{s/t}_{\kv,\kv'}\frac{\Delta^{s/t}_{\kv'}}{2E^{s/t}_{\kv'}}\tanh\Big(\frac{E^{s/t}_{\kv'}}{2k_BT}\Big),\nonumber\\
&&\label{eq:SCGapEquationNS}
\end{eqnarray}
where
\begin{eqnarray}
E_\kv&=&\sqrt{\xi_{\kv}^2+|\Delta^{s/t}_\kv|^2}
\end{eqnarray}
are the energy eigenvalues in the superconducting state with singlet/triplet order parameter $\Delta^{s/t}_\kv$ and $\Gamma^{s/t}_{\kv,\kv'}$ are the (effective) singlet/triplet pairing interactions.

The pairing vertices are given by the symmetrized opposite spin pairing vertices for the singlet case
\begin{equation}
\Gamma^{s}_{\kv,\kv'}=\Gamma^{\rm opp. sp}_{\kv,\kv'}+\Gamma^{\rm opp. sp}_{-\kv,\kv'},
\end{equation}
and by the same spin pairing vertices for the triplet case which are also related to the opposite spin pairing vertices as follows
\begin{equation}
\Gamma^{t}_{\kv,\kv'}=\Gamma^{\rm opp. sp}_{\kv,\kv'}-\Gamma^{\rm opp. sp}_{-\kv,\kv'}=\Gamma^{\rm same~sp}_{\kv,\kv'}-\Gamma^{\rm same~sp}_{-\kv,\kv'}.
\label{eq_gammat}
\end{equation}
Note that the singlet (triplet) pairing vertices are symmetrized (antisymmetrized) to obey the Pauli principle for the overall antisymmetry of the Cooper pair wavefunction. For the singlet case this is apparent in the spin structure, while a triplet state must be of odd parity in the order parameter, $\Delta_{\kv}^s=\Delta_{-\kv}^s$ and $\Delta_\kv^t=-\Delta_{-\kv}^t$.

If the pairing interaction is mediated by a spin-fluctuation mechanism, the interactions can be approximately derived from a (partial) summation of bubble and ladder diagrams with the result~\cite{Scalapino1986}
\begin{eqnarray}
 \Gamma_{\kv,\kv'}^\textrm{opp.sp}&=&U+\frac{U^2}{2}\chi_{(\kv-\kv')}^{sp}-\frac{U^2}{2}\chi_{(\kv-\kv')}^{ch}+U^2\chi_{(\kv+\kv')}^{sp},  \label{eq:PMoppspin} \nonumber\\
&& \\
 \Gamma_{\kv,\kv'}^\textrm{same sp}&=&-\frac{U^2}{2}\left (\chi_{(\kv-\kv')}^{sp}+\chi_{(\kv-\kv')}^{ch}\right),  \label{eq:PMsamespin}
 \end{eqnarray}
 where the spin and charge susceptibilities are given by
  \begin{eqnarray}
 \chi^{sp}_\qv&=&\frac{\chi_0(\qv)}{1-{U}\chi_0(\qv)}, \label{eq:chisp}\\
 \chi^{ch}_\qv&=&\frac{\chi_0(\qv)}{1+{U}\chi_0(\qv)}\label{eq:chich}
\end{eqnarray}
in the random phase approximation (RPA).

Here, $U$ is an (effective) Hubbard interaction originating from the Hamiltonian
\begin{align}
 H_{\text{int}}=\frac{U}{2N}\sum_{\kv,\kv',\qv}\sum_{\spin}c_{\kv'\spin}^\dagger c_{-\kv'+\qv\ospin}^\dagger c_{-\kv+\qv\ospin} c_{\kv\spin}
\end{align}
and $\chi_0(\qv)$ is the response function (susceptibility) of the corresponding noninteracting electron system.
In principle, one would like to calculate $\chi_0(\qv)$   directly from the tight binding model Hamiltonian (\ref{eq_tb}) and this has been done previously  in various theoretical investigations in single band systems\cite{Romer2015} and multi band systems\cite{Eremin2002,Kuroki2008,Graser2009}.  Note however that while the pairing problem is a low energy problem where only the electronic structure very near the Fermi level plays a role, in the multiband case,  the pairing interaction itself may in addition contain contributions involving interband transitions from higher energy states. The paramagnetic susceptibility in itinerant multiband models is given by an integral over the Brillouin zone and sum over bands, where all possible scattering contributions are taken into account, weighted by the inverse energy difference of the eigenstates. Thus, for multiband models, interband transitions can contribute significantly if there is sufficient phase space connecting states with the same momentum transfer $\qv$. The importance of these scattering contributions has been examined using the Kramers-Kroning relation for example using a model for an iron-based material in Ref. \cite{Kreisel15a}.

In addition, as we discuss further below in Section \ref{sec:ute2}, the Fermi surface of many interesting candidate unconventional superconductors is not well known.  Since many such systems are strongly correlated materials, density functional theories have well known limitations; the band structure of heavy fermion systems, in particular, is notoriously difficult to calculate accurately due to their significant $f$-electron character at the Fermi level.  While comparison with angle-resolved photoemission spectroscopy (ARPES) and quantum oscillation experiments sometimes assists these determinations, in many cases atomically flat surfaces appropriate for ARPES are difficult to obtain, and the 3D nature of the systems introduces further uncertainties. 
We therefore adopt a slightly different viewpoint in considering the low energy tight binding model  Eq. (\ref{eq_tb}) and the response function $\chi_0(\qv)$ as independent quantities, an approach that has been adapted already by calculating $\chi_0(\qv)$ from more bands than were used in the solution of  the actual pairing problem\cite{Ikeda2016}.

In the following, we restrict ourselves further to the superconducting instability at $T_c$.
Using an expansion in powers of the gap, the BCS gap equation, Eq. (\ref{eq:SCGapEquationNS}) can be cast into the linearized gap equation in the singlet and triplet channel
\begin{equation}
\Big[ -\frac{1}{2(2\pi)^D}\int_{FS}\frac{d\kv'}{|v_{\kv'}|}\Gamma^{s/t}_{\kv,\kv'}\Big]g({\kv'})= \lambda_i g(\kv).
\label{eq:lge}
\end{equation}
Here $g(\kv)$ is the gap symmetry function which contains the momentum dependence of the order parameter and $\lambda_i$ is the eigenvalue. The leading instability is given by the largest eigenvalue and $g(\kv)$ is proportional to the gap function at $T_c$. In this way, the calculation reduces to the calculation of the eigenvalues and eigenvectors of a matrix whose elements need to be determined by calculating the pairing interaction and the weight associated with each of the k-points that parametrize the Fermi surface in $D=2,3$ dimensions.

\section{Results for model systems}

\subsection{Triplet pairing from scattering at finite momentum}
\begin{figure}[tb]
     \includegraphics[width=\linewidth]{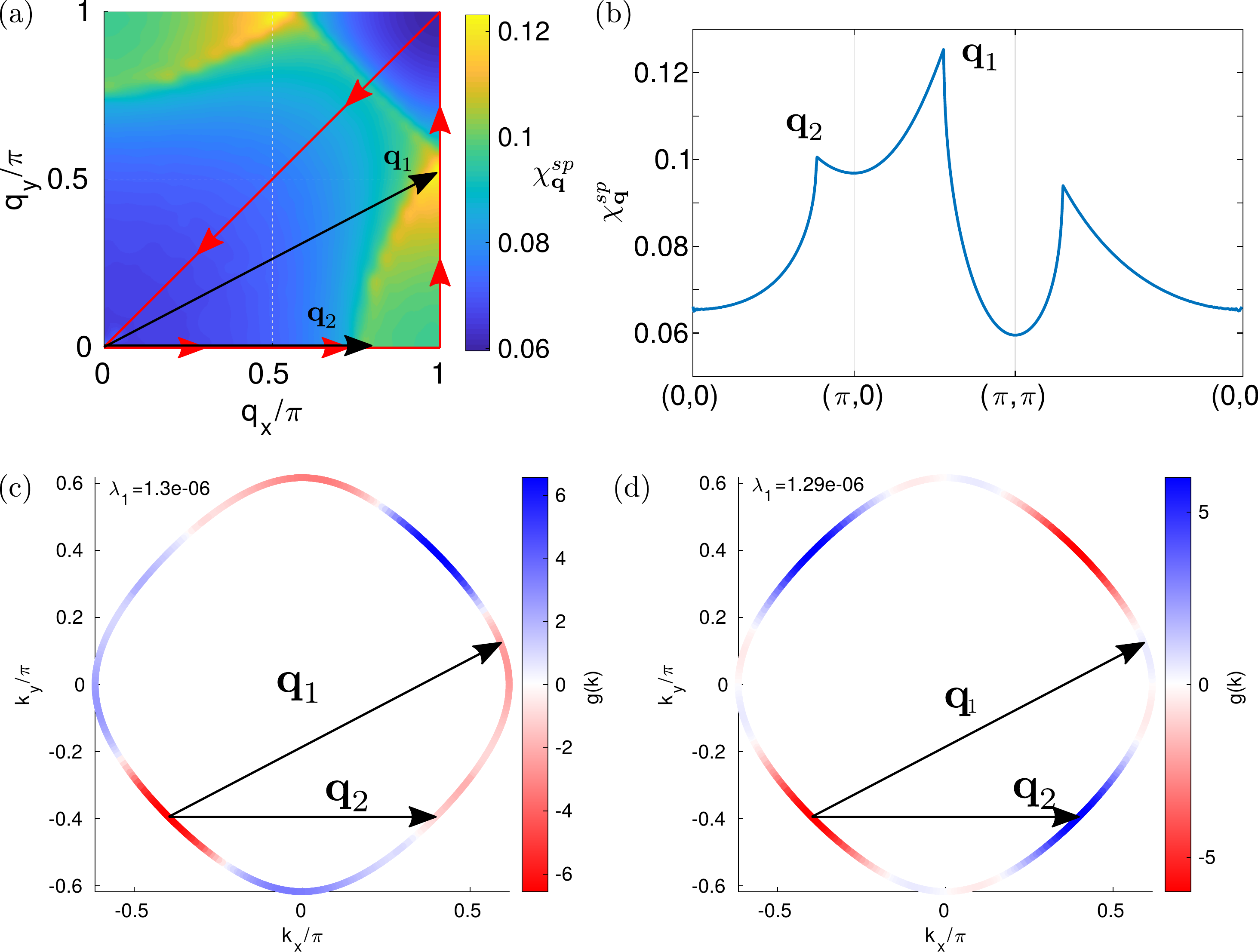}
 \caption{Single band Hubbard model: Triplet pairing from finite momentum scattering (a) Spin susceptibility in one quarter of the Brillouin zone for the single band Hubbard model ($U=0.05$, $t=1$, $t'=0$) at filling $n=0.54$ and low temperature $T=0.0001$ which exhibits two dominant peaks at the scattering vectors $\qv_1$ and $\qv_2$ as explicitly shown in panel (b) where the susceptibility is plotted along the high symmetry path marked in red in panel (a). (c) Gap symmetry function $g(\kv)$ of the leading triplet instability and (d) leading singlet instability plotted at the Fermi surface with the same momentum scale as in (a) and the scattering vectors placed to connect points $\kv$ and $\kv'$ on the Fermi surface.}
    \label{Fig_single_band}
\end{figure}
Examining the spin singlet pairing vertex, one finds that $\Gamma^{s}_{\kv,\kv'}>0$ for all choices of $\kv$ and $\kv'$ when assuming  $\chi^{sp}_\qv> \chi^{ch}_\qv$ which is natural in a RPA approach, see Eqs. (\ref{eq:chisp}), (\ref{eq:chich}). Thus  the spin singlet pairing is always repulsive for pairs of states at the Fermi surface,
such that only sign changing solutions to the gap equation, Eq. (\ref{eq:SCGapEquationNS}) are allowed; the additional minus sign is then canceled by the relative minus sign in $\Delta_\kv$ and $\Delta_{\kv'}$. Indeed, it is required that the pairing vertex is actually momentum dependent, originating from a momentum dependent susceptibility $\chi_0(\qv)$ mediating a stronger repulsion for some $\kv+\kv'=\qv$ or $\kv-\kv'=\qv$. In this case the gap symmetry function $g(\kv)$ tends to be large and of {\it opposite sign} for points $\kv$ and $\kv'$ on the Fermi surface that are connected by the vector $\qv$ where $\chi_0(\qv)$ is peaked.

For the triplet pairing vertex, it turns out that $\Gamma^{t}_{\kv,\kv'}$ in Eq. (\ref{eq_gammat}) is a difference of two positive terms and can be positive or negative, depending on whether $\Gamma^{\rm same~sp}_{\kv,\kv'}$ or $\Gamma^{\rm same~sp}_{-\kv,\kv'}$ is larger. Furthermore, we note that these do only depend on the momentum transfer $\kv-\kv'$, see Eq. (\ref{eq:PMsamespin}).  If the summed charge and spin susceptibilities at $\qv=\kv-\kv'$ are larger (than the ones at $\kv+\kv'$), one has an effective attractive interaction for Cooper pairs $(\kv,-\kv)$, $(\kv',-\kv')$ at the respective Fermi points $\kv$ and $\kv'$. Indeed, a susceptibility that is peaked at $\qv=0$ would immediately give rise to an attractive pair scattering for Cooper pairs that are located close by on the Fermi surface.

To find a solution to the gap equation Eq. (\ref{eq:SCGapEquationNS}), it is required  additionally that the order parameter is odd parity, yielding a sign change upon inversion. In addition, the periodicity of the gap function within the Brillouin zone  forces a vanishing $g(\kv)$ at the zone boundaries. Examining the single band Hubbard model alone, one finds triplet instabilities to be leading solutions for small filling where the gap symmetry function is of lowest order harmonic, i.e. has the form $\propto \sin(k_x)$ in two dimensions, see for example Refs. \cite{Romer2015,Kreisel2017_JSNM}.

If, however, the susceptibility is  peaked at $\qv$ with magnitude slightly smaller than $2k_F$, there are further solutions to the linearized gap equation, Eq. (\ref{eq:lge}) which involve higher- order harmonics, i.e. more nodal lines/points are required. Such a situation for example occurs in the single band model close to quarter filling \cite{Romer2015,deng15} where in the weak coupling regime $U\ll t$ a triplet state  emerging from finite $\qv$ scattering becomes favorable and has an eigenvalue that exceeds the one from the singlet instablility. For purposes of illustration,  we show the  susceptibility for this case in Fig. \ref{Fig_single_band} (a,b) together with the leading instabilities in the triplet and singlet channel (c,d). Note that the position of accidental nodal lines and those where $|g(\kv)|$ is maximum is very similar in these states.
As seen in panels (a,b), the susceptibility has two dominant peak contributions labeled by $\qv_1$ and $\qv_2$. These mediate the dominant pairing interaction that drives $g(\kv)$ to be large and the {\it same sign} for the triplet case because $\Gamma^{t}_{\kv,\kv'}<0$ (c), while in the singlet case the same scattering vectors connect parts of the Fermi surface which have opposite sign of the order parameter.

\subsection{Phase diagram from peaked susceptibility}

In this section, we drop the requirement that the (bare) susceptibility is directly calculated from the tight binding model that describes the Fermi surface and therefore the low-energy electronic structure of the system. Instead, the paramagnetic response is assumed to exhibit a peak structure at a certain dominant vector $\qv_0$ (and its symmetry related counterparts); the origin of $\chi_0(\qv)$ could be from higher energy scattering (other bands) or be related to a Heisenberg model of localized spins with dominant fluctuations at the symmetry-related $\qv_0$.
In the localized picture, the momentum of the dominant fluctuations is given by the
structure of the exchange interactions in the model such that in principle any $\qv_0$ can
be realized with a suitable microscopic model.

\begin{figure}[tb]
  \includegraphics[width=\linewidth]{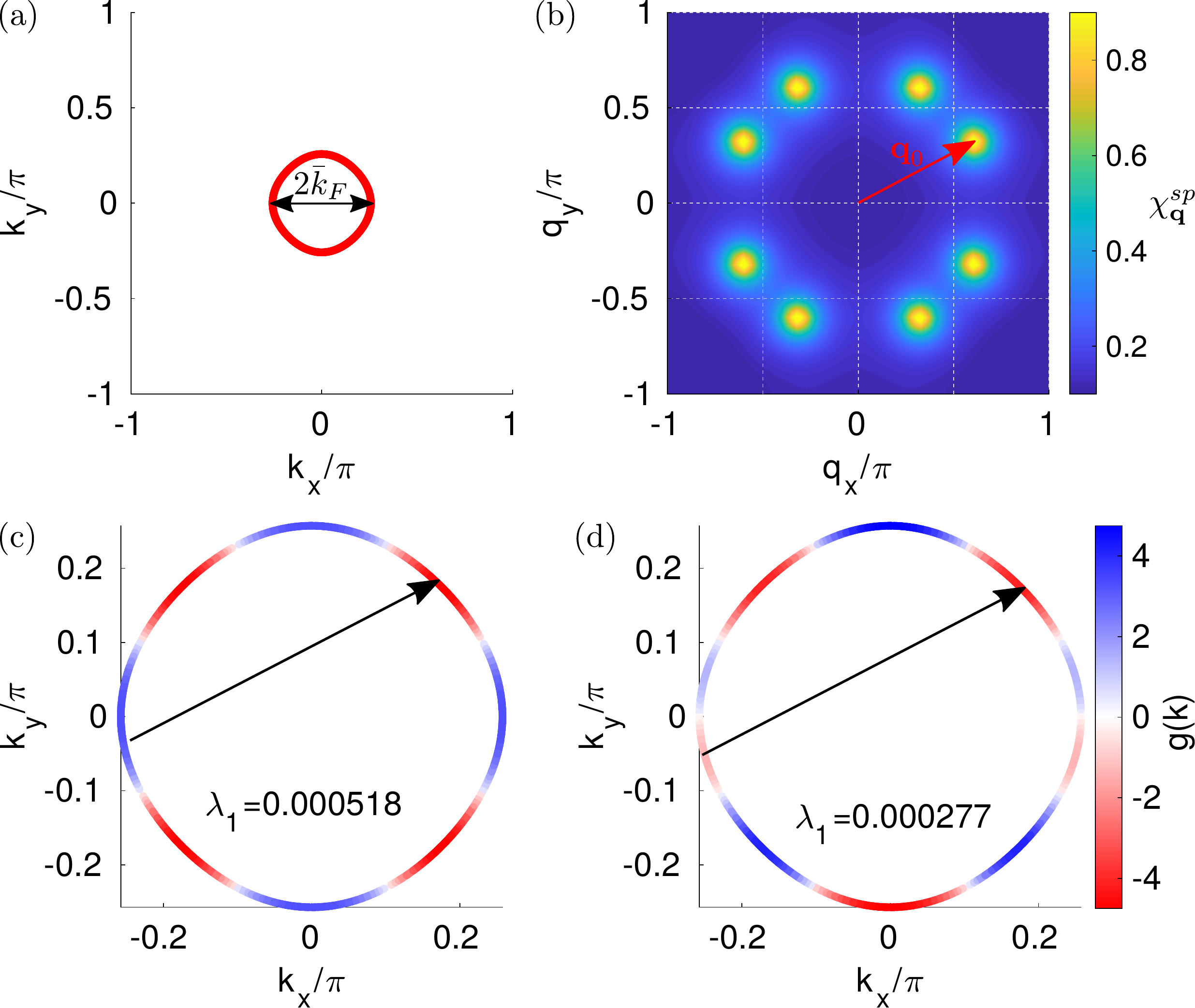}
 \caption{Single band model with phenomenological susceptibility (a) Fermi surface of the model at filling $n=0.1$ exhibiting an almost circular Fermi surface with average magnitude $\bar{k}_F$ of the Fermi vector. (b) Susceptibility exhibiting peaks at $\qv_0$ and symmetry related points in the Brioullin zone. (c,d) singlet and triplet gap structure functions $g(k)$ for the respective leading solution in each channel.}
 \label{Fig_chi_single_band}
\end{figure}
For concreteness, we consider our tight binding model on the square lattice with $t=1$, $t'=-0.2$ at filling $n=0.1$, i.e. $\mu=-2.82877$,  yielding an almost circular Fermi surface, see Fig. \ref{Fig_chi_single_band} (a). An example susceptibility with peaks at $\qv_0=(q_x,q_y)=(1.9,1)$ and symmetry related positions for the D$_4$ symmetry is presented in panel (b) of the same figure. The peak is taken to have a Lorenzian shape 
with a structure $f_{\qv_0}(\qv)=1/[(\qv_0-\qv)^2/\zeta^2 +1]$ where we set the decay constant in momentum space as $\zeta=0.5$. In summary, we use
\begin{align}
 \chi_0(\qv)=\frac 1 {N_0} \sum_{D_4/\qv_0} f_{\qv_0}(\qv) \label{eq_chi_phen}
\end{align}
where $N_0$ is chosen such that the maximum (in the Brillouin zone) is set to unity, i.e. the critical Hubbard interaction would be at $U_c=t=1$ and the sum is taken over the 8 symmetry related vectors.
Using this susceptibility together with the pairing vertex, Eqs. (\ref{eq:PMoppspin}), (\ref{eq:PMsamespin}), (\ref{eq:chisp}) and (\ref{eq:chich})  to solve the linearized gap equation for $U=0.9$   yields an (extended) s-wave state Fig. \ref{Fig_chi_single_band} (c) and a (higher order) p-wave triplet state that has lower eigenvalue (but of similar order) (d). Note that the dominant scattering processes connect Fermi points with the opposite (same) sign for the singlet (triplet) states. Because the scattering vector $\qv_0$ is not at a high symmetry point in general, the triplet state can retain odd parity, as seen in Fig. \ref{Fig_chi_single_band}(d).

\begin{figure}[tb]
  \includegraphics[width=\linewidth]{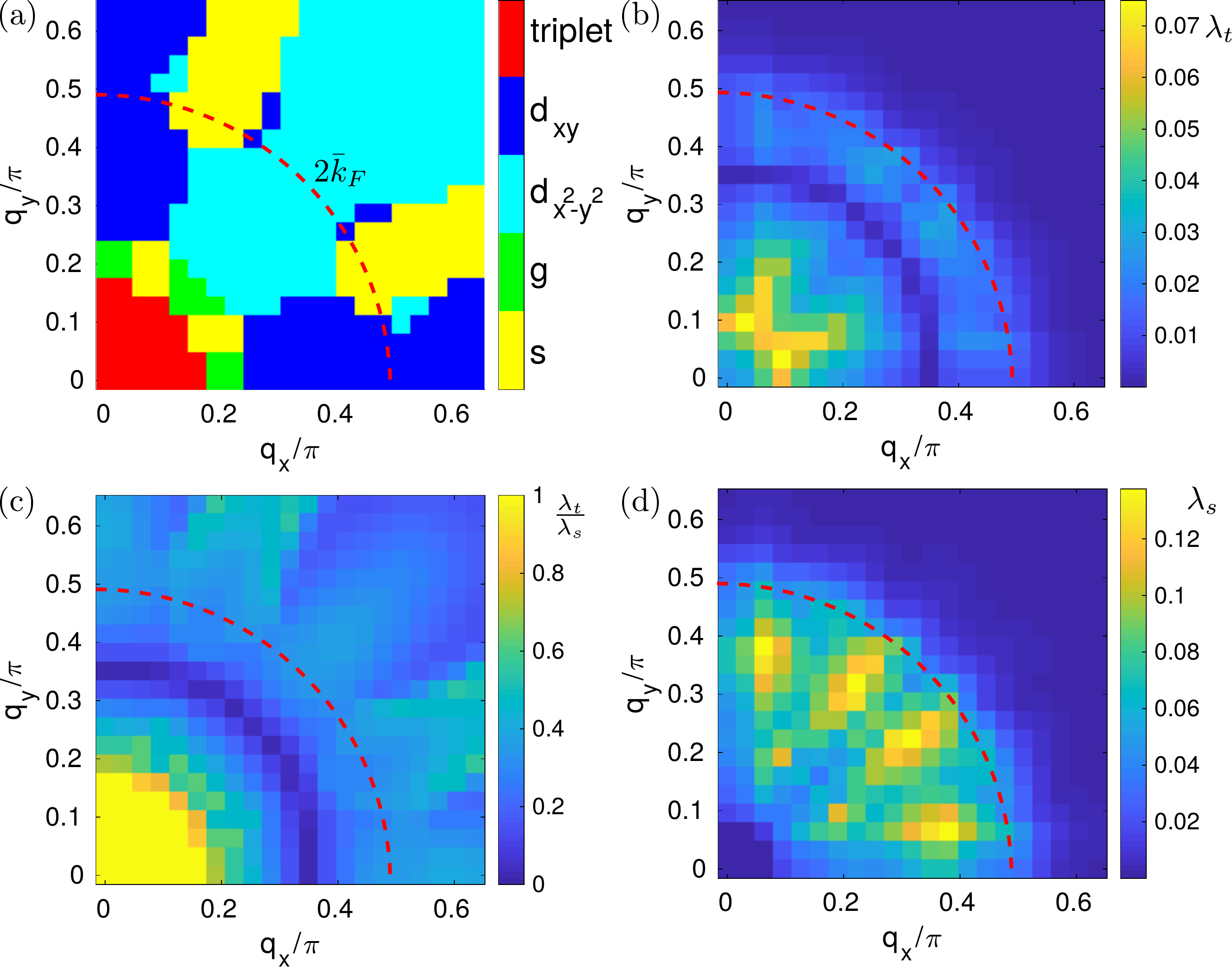}
 \caption{Phase diagram of the single band model (a) Phase diagram of dominating superconducting instabilities as a function of the position of the peak in the susceptibility $\qv_0=(q_x,q_y)$, colors indicate the types of solutions. (b,d) Eigenvalue $\lambda_t$ ($\lambda_s$) of the leading triplet (singlet) solution. The triplet eigenvalue
 is peaked at small, but nonzero momentum transfer, while the singlet eigenvalue has a slightly larger maximal eigenvalue and strong peaks closer to the $2\bar{k}_F$ line. (c) The ratio of the eigenvalues $\lambda_t/\lambda_s$ showing the relative competition which is larger than 1 on a finite circle in the center, has a dip and a rise beyond the dashed $2\bar{k}_F$ line.}
 \label{Fig_phase_diagram_single_band}
\end{figure}

Next, we discuss the phase diagram as a function of the peak position $\qv_0$, which exhibits various dominant instabilities as presented in Fig. \ref{Fig_phase_diagram_single_band} (a), notably the finite size of the triplet phase for small $|\qv_0|$. The phase diagram is only shown for $\qv_0$ in the first quadrant and there is a mirror symmetry at the 45 degree line as expected from the square lattice system. Panels (b) and (d) show the eigenvalue of the leading instability in each channel, while panel (c) presenting the relative eigenvalues $\lambda_t/\lambda_s$ reveals that a triplet state can become competitive also for large $\qv_0$ close to phase boundaries and ''triple`` points where three singlet instabilities meet, very similar to the emergence of the dome for triplet states in the single band Hubbard model at weak coupling\cite{Romer2015,deng15}.
In this example, however, the triplet state is never the leading instability except near $\qv=0$.

\begin{figure}[tb]
  \includegraphics[width=\linewidth]{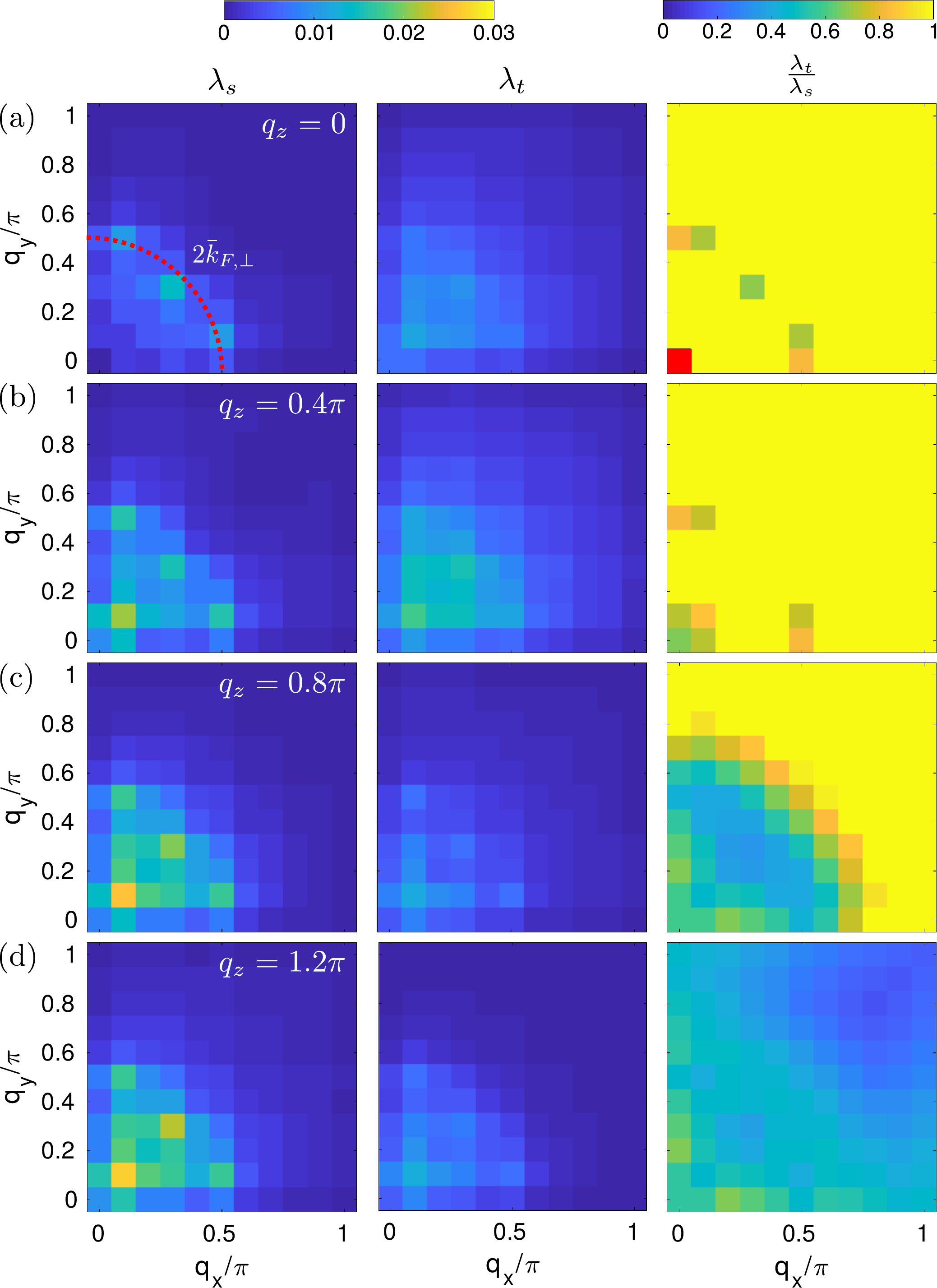}
 \caption{Singlet pairing eigenvalues $\lambda_s$ and triplet pairing eigenvalues $\lambda_t$ as function of $\qv_0=(q_x,q_y,q_z)$ for a single band model at low filling $n=0.1$, but with trivially extended dispersion in the z-direction (a-d). At small $q_z$, triplet solutions are competitive (a,b), while for larger $q_z$, the phase diagram is dominated by singlet instabilities (c,d). The line of twice the average in plane Fermi momentum $2\bar{k}_F$ is marked in panel (a).}
 \label{Fig_chi_toy_3d_phase}
\end{figure}

\begin{figure}[tb]
  \includegraphics[width=\linewidth]{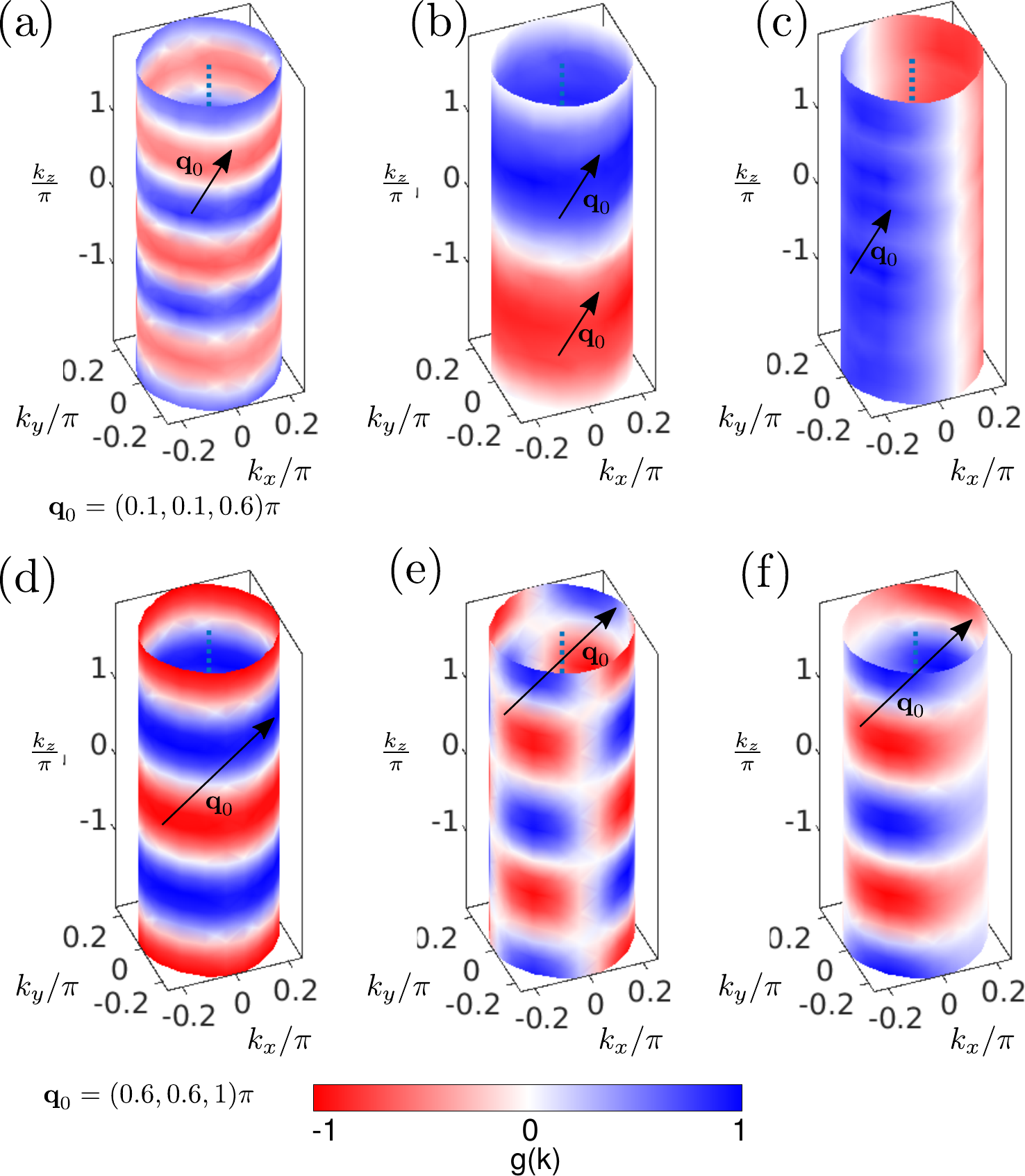}
 \caption{Examples of pairing states for a model without dispersion in the z-direction, but three dimensional structure of the susceptibility. (a) Leading singlet state for $\qv_0=(0.1,0.1,0.6)\pi$ where accidental nodes appear in the $k_x-k_y$ planes. (b,c) The two leading triplet states of $p_z$ symmetry and $p_x$ symmetry where the pair scattering is within the region of the same sign. (d-f) Pairing states for a different $\qv_0=(0.6,0.6,1.0)\pi$  (d-e) The leading singlet states shows nodes at $k_x-k_y$ planes. (f) The leading triplet solution is of higher order harmonics with $\qv_0$ connecting same sign parts across a nodal line.}
 \label{Fig_examples_pairing_simple_3d}
\end{figure}

Extending the normal state Hamiltonian trivially in the $k_z$ direction (with no dispersion), the Fermi surface becomes a cylindrical tube and we can study the influence of finite $q_z$ components in the susceptibility. Indeed for $q_z=0$, our model just consists of sets of coupled Cooper pairs at the Fermi points $\kv$ and $\kv'$ at each value of $k_z$, while there is no scattering between states at different $k_z$. In this case we will just recover highly degenerate solutions that have pairing in each plane. Since we employ only roughly 650 k-points on the Fermi surface, the numerical approach does not faithfully capture these degenerancies. However at finite $q_z$ one finds competition between singlet and triplet instabilities, Fig. \ref{Fig_chi_toy_3d_phase} (c,d), with triplet states that have a nodal plane at $k_z=0$ (Fig. \ref{Fig_examples_pairing_simple_3d} (b) or states where (accidental) nodes appear across the scattering vector $\qv_0$, see Fig. \ref{Fig_examples_pairing_simple_3d} (e,f).
Such triplet states with additional (non-symmetry-enforced) nodes have been discussed in the context of spin fluctuation pairing in Sr$_2$RuO$_4$\cite{Raghu2010a,Scaffidi2014}.
The calculations and discussions in the previous section demonstrate that pairing into a triplet state can be driven by spin fluctuations at small $\qv_0$, where the pair scattering is trivially between states that are in a region with the same sign of the superconducting order parameter, or by spin fluctuations with large momentum transfer $\qv_0$ where this vector connects regions of the Fermi surface that exhibit the same sign, but have two or more (accidental) nodal lines in between.

\section{Discussion of relation to \texorpdfstring{UTe$_2$}{UTe2}}
\label{sec:ute2}

In this section, we discuss the electronic structure of the candidate triplet superconductor UTe$_2$ and use a similar phenomenological approach to motivate possible superconducting instabilities assuming the Fermi surface of this system as revealed from ab-initio calculations\cite{Dai2020,Miao2020} and ARPES measurements\cite{Miao2020}.
We begin by studying a featureless Fermi surface in a single band picture, an approximation that neglects multi-orbital effects and more importantly spin-orbit coupling when calculating the pairing interaction that enters Eq. (\ref{eq:lge}). Nevertheless, the Fermi surface itself is derived from a Hamiltonian with SOC, such that the pairing may be supposed to be in the pseudospin basis of the Kramer's degenerate states on the Fermi surface.

However, for the sake of discussing the possibility of instabilities from nonzero $\qv_0$ fluctuations, our approach is simplified, but it captures the essential physics of pairing from antiferromagnetic fluctuations. We note that by adopting a multiband BCS-like approach, without including the frequency dependent self-energy as in Eliashberg theory, we neglect band renormalizations that may be stronger in triplet superconductors and destabilize the triplet channel further\cite{Arita2000}.  We believe that this quantitative effect will simply shift the phase boundaries of the triplet phases found below slightly, and will not affect our main qualitative point here.

Let us discuss at this point the influence of anisotropy in spin space, as in principle expected when spin-orbit coupling is present. This has been examined by Kuwabara and Ogata\cite{Ogata2000}, 
where the 
 susceptibility obtained from a calculation using a DFT band structure for Sr$_2$RuO$_4$ (fixing the maximum of the spin correlations and allowing for a spin anisotropy with  $\chi_{zz}> \chi_{+-}$) was  discussed in terms of its influence on  superconducting pairing. Decreasing $\chi_{+-}$ leaves the triplet instabilities with a $d$-vector in the $x-y$ plane unchanged, while the pairing for the state with a $d$-vector along the $z$ direction is modified, as is the singlet pairing interaction. For their choice of the Fermi surface topology with a nesting across the Brillouin zone boundary it was demonstrated that a lowest harmonic d-wave singlet solution is suppressed upon decreasing $\chi_{+-}$ and the aforementioned triplet solution ($d$-vector along $z$ direction) of the lowest harmonic is enhanced such that their $T_c$'s cross, implying a phase transition from singlet to triplet pairing states. Indeed, for UTe$_2$ substantial anisotropy in the spin susceptibility has been detected  \footnote{P. Dai, private communication}, such that splitting of $T_c$ for the triplet states with different directions of the $d$-vector is expected as well. However, whether a reduced $\chi_{+-}$ will make the triplet state with a $d$-vector the along $z$ direction more (less) competitive depends on whether the leading superconducting state is driven by scattering between $\kv$ and $\kv'$ which have the opposite (same) sign of $\Delta(\kv)$. In the aforementioned work, instabilities with lowest order harmonics are considered, such that the possibility of driving a triplet state by fluctuations at $\qv$ connecting the same sign of the order parameter with higher order nodes (see Fig. \ref{Fig_single_band} (d)) is not included. This always leads to a negative expectation value for the pairing eigenvalue when calculated without anisotropy in spin space as discussed in Ref. \onlinecite{Ogata2000}.

\subsection{Electronic structure of \texorpdfstring{UTe$_2$}{UTe2}}
 \begin{figure*}[tb]
  \includegraphics[width=\linewidth]{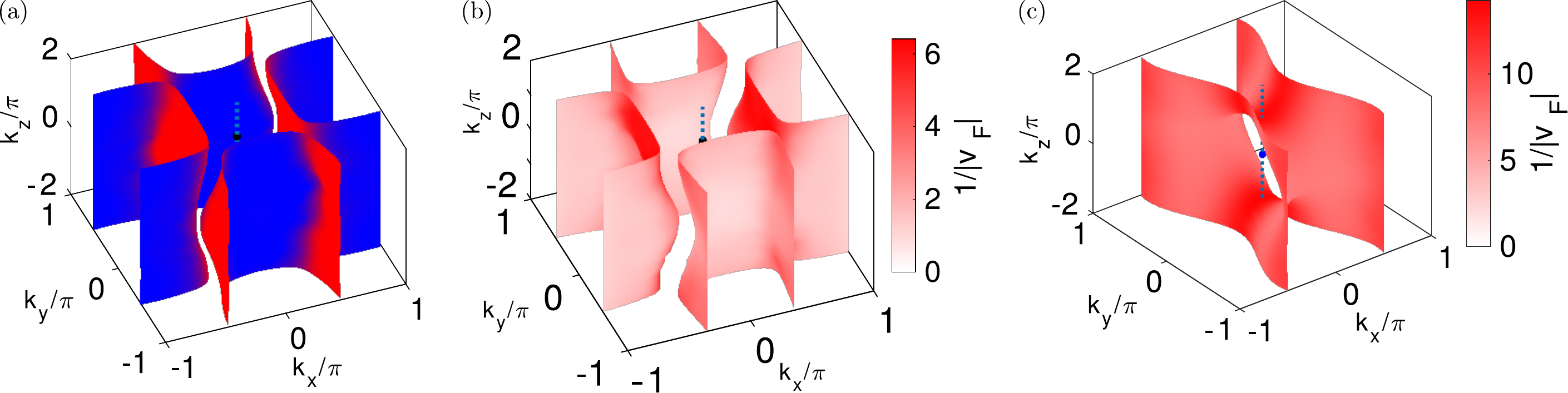}
 \caption{Fermi surfaces from theoretical models. (a) Fermi surface from a 4 band tight binding model within a ab initio calculation for ThTe$_2$ (red/blue) different orbital content (this work). (b) Fermi surface from a 56 band tight binding model derived from an ab initio calculation for UTe$_2$ (shown is the inverse Fermi velocity in units of 1/eV, this work). (c) Fermi surface from a single band model fitted to ab initio calculations \cite{Shishidou2021}.}
 \label{Fig_fermi_UTe2}
\end{figure*}

Next, we want to understand which specific materials characteristics  might
drive triplet superconductivity UTe$_2$,
which shows considerable experimental evidence 
counter to singlet pairing. At the same time, to our knowledge there is no direct evidence for the expected ferromagnetic spin fluctuations (despite the claimed ``proximity"  to other sister compounds with ferromagnetic order)\cite{Aoki_review2019}.
Instead, antiferromagnetic spin fluctuations were detected in inelastic neutron scattering experiments\cite{Dai2020,duan2021resonance,raymond2021feedback}. While it is challenging to obtain a realistic electronic structure
(tight binding model) from ab-initio calculations because the system is a heavy fermion material with strong spin-orbit coupling effects and electronic correlations,  there have been
several attempts presented so far. Some authors have adopted
calculations of the electronic structure that simply study the sister compound ThTe$_2$ with the experimental lattice constants of UTe$_2$, arguing that the primary $f$-weight is away from the Fermi level \cite{Miao2020}, a detail that is also shown in the supplementary material. Such an approach evidently neglects low-energy  Kondo lattice scattering.   
Nevertheless, it appears
that at least some of  the effects of correlations
are captured within this approximation, as confirmed by DMFT calculations using the correct ab-initio input for UTe$_2$\cite{Miao2020}, as well as DFT+U calculations\cite{Shick2019}.  Here it is seen that the UTe$_2$ DMFT  spectral function is quite similar to a ThTe$_2$ calculation and compares reasonably well to ARPES measurements. From such an approach, one can derive a 4 band model with two U atoms in the elementary cell, exhibiting a Fermi surface as presented in Fig. \ref{Fig_fermi_UTe2} (a).

We have obtained the corresponding tight binding model from a DFT calculation with the FPLO code\cite{PhysRevB.59.1743} using the crystallographic parameters of UTe$_2$ with $a=4.1617$\AA{}, $b=6.1276$\AA{} and $c=13.965$\AA{}, the space group Immm (\#71) and internal parameters for Th at Wyckoff
position 4e $z=0.1348$, Te at 4j $z=0.2977$ and Te at 4h $y=0.251$.

A very similar Fermi surface with 4 almost two dimensional, nearly square tubes, but somewhat different corrugation in the $k_z$ direction is obtained from a tight binding model derived from a LDA+$U$ calculation for UTe$_2$ (see Fig. \ref{Fig_fermi_UTe2} (b)), with the same lattice parameters as described in the previous paragraph, but placing an U atom.
Fully relativistic calculations were carried out, which treats spin-orbit coupling accurately.
Here, we use the values $F_0 = 2 eV$, $F_2 = 0$ and $F_4 = 0$ for the Slater parameters corresponding to $U_{\mathrm{eff}} = 2eV$, $J = 0$ and perform a DFT+U calculation. The double counting term in LDA+U calculations is treated using the
around mean field (AMF) formula.\cite{PhysRevB.79.035103}
After initial convergence, we project to symmetry preserving Wannier functions of 56 states near the Fermi level\footnote{Details in the supplementary material which includes the references \cite{PhysRevB.60.14035,PhysRevB.59.1743,PhysRevB.45.13244,PhysRevB.103.125136,doi:10.1143/JPSJS.75S.116,Miao2020,Ishizuka2019}}

Finally, a somewhat different topology of the Fermi surface was recently proposed by a fit of a single band model (including spin-orbit coupling) to a DFT+$U$ calculation, (see Fig. \ref{Fig_fermi_UTe2} (b))\cite{Shishidou2021}. The main differences are whether or whether not the bands cross the Fermi level along the dashed line from $\Gamma$ to the top of the Brillouin zone, an effect that theoretically can be controlled by the choice of the Hubbard interaction U in the DFT+$U$ approach such that band crossings and band inversion of bands with even and odd parity can be induced\cite{Shishidou2021}.

\subsection{Susceptibility and spin fluctuation pairing on the Fermi surface of \texorpdfstring{UTe$_2$}{UTe2}}
 \begin{figure}[tb]
  \includegraphics[width=\linewidth]{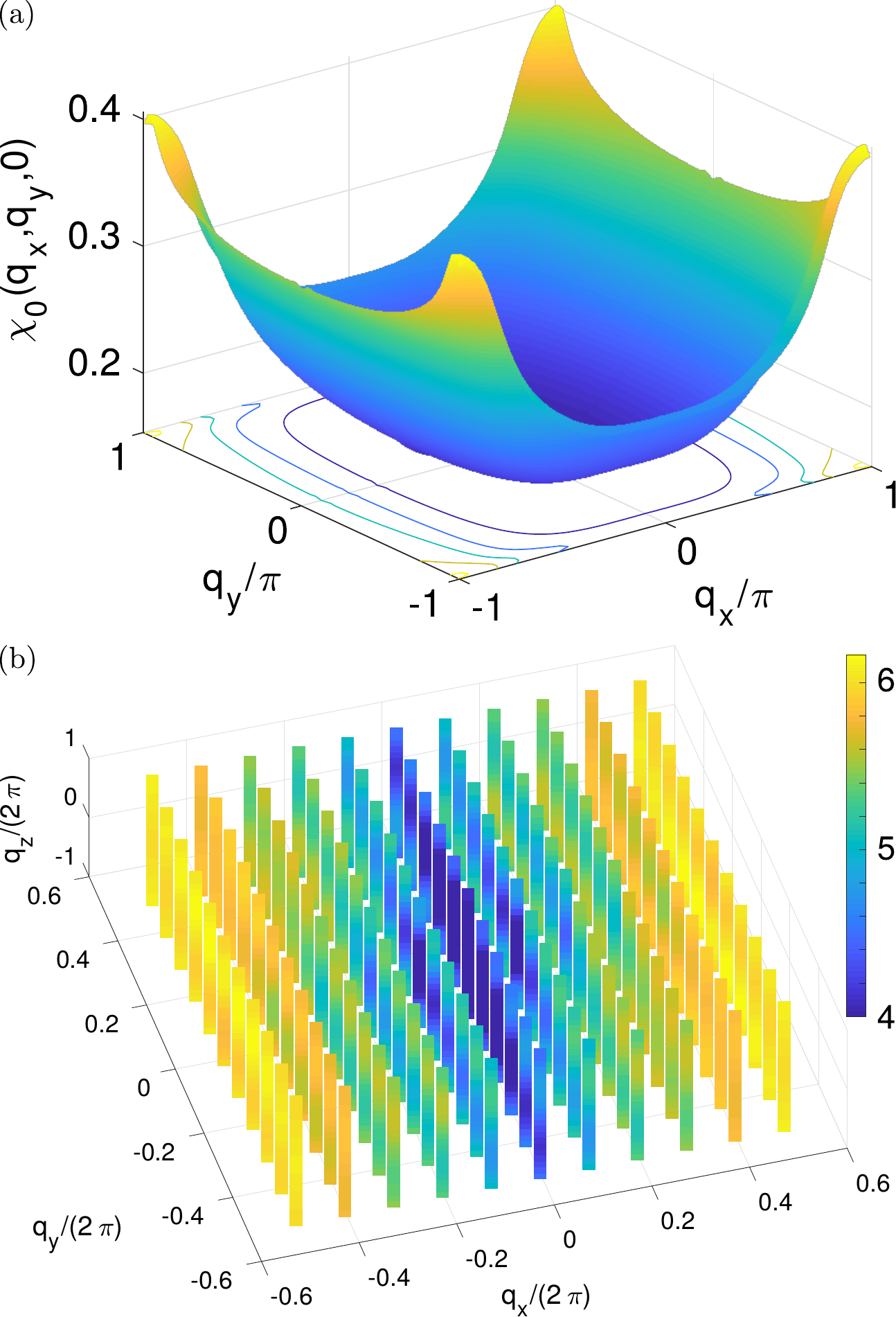}
 \caption{Physical susceptibility (in units $1/\mathrm{eV}$) from tight binding models. (a) Susceptibility as calculated from the 4 band model (at $q_z=0$) and (b) visualization of the susceptibility
 from the 56 band model in the full Brillouin zone with totally 11$\times$11$\times$21 $\qv$-points repeating the values at the Brillouin zone boundary. The color of each pixel corresponds to one data point; the data for fixed $q_x$ and $q_y$ appear as stacked bars in $q_z$ direction. Units for $\chi(\qv)$ are $1/eV$, but are not comparable due to differences in the overall electron density of the models.}
 \label{Fig_chi_UTe2}
\end{figure}
For our 56 band tight binding model, we calculate the orbitally resolved susceptibility\cite{Graser2009} 
\begin{align}
\chi_{\ell_1 \ell_2 \ell_3 \ell_4}^0 (\qv) &= -\frac{1}{N} \sum_{\kv,\mu \nu} \left[ f(E_\nu (\kv+\qv)) -f(E_\mu (\kv)) \right]\nonumber \\
& \times 
\frac{a_\mu^{\ell_4}(\kv) a_\mu^{\ell_2 *}(\kv) a_\nu^{\ell_1}(\kv+\qv) a_\nu^{\ell_3 *}(\kv+\qv)}{
E_\nu (\kv+\qv) - E_\mu(\kv)} 
\label{eq:spinsusceptibility}
\end{align}
for the components $\ell_1=\ell_3$, $\ell_2=\ell_4$ which enter the physical susceptibility
\begin{align}
 \chi_0(\qv)=\frac 12 \sum_{\ell,\ell'} \chi_{\ell \ell' \ell \ell'}^0 (\qv)\,.
\end{align}
Here, $a_\mu^{\ell}(\kv)$ is the $\ell$ orbital component of the eigenvector in band $\mu$ with eigenenergy $E_\mu (\kv)$ and $f(E)$ the Fermi function.
We calculate the susceptibility on a grid consisting in $11 \times 11\times 21$ equally spaced momenta $\qv$ in the Brillouin zone $[-\pi/a \ldots \pi/a]\times [-\pi/b \ldots \pi/b]\times [-2\pi/c \ldots 2\pi/c]$ (Fig. \ref{Fig_chi_UTe2}(b)), where the sum over bands $\mu$ and $\nu$
is cut out when large band energies $E_{\mu}$ appear in the denominator.
For the 4 band model, we can use a much finer $\kv$ grid and we do not have to cut off any bands from the sum.
 Comparing the susceptibility as obtained from the two tight binding models reveals that there is no signature of ferromagnetic fluctuations since $\chi(\qv)$ has a minimum at $\qv=0$, but a ridge close to the Brillouin zone boundary in x-direction, see Fig. \ref{Fig_chi_UTe2}. There are small variations as function of $q_y$ and $q_z$ in both cases which might play a role in the formation of the superconducting instability.
Note that a pairing calculation within a 
spin-fluctuation approach
for the 4 band model yields a leading extended s-wave state with nodes at $k_x$-$k_y$ planes while the triplet state has much smaller eigenvalue (not shown). The ab initio calculation of tight binding models for UTe$_2$ might be inaccurate, and our spin-fluctuation approach does not incorporate effects of spin-orbit coupling in the 4 band model. We therefore do not advocate for a realistic determination of the pairing state from these investigations. Instead, we want to follow up on the approach to use the example Fermi surface geometry in order to investigate how spin-fluctuation pairing would lead to singlet and triplet instabilities given a (paramagnetic) susceptibility with a given peak structure. Guided by the  existence of the ridge of intensity in the $\chi(\qv)$ in both models and also experimental findings of large antiferromagnetic fluctuations close to the Brillouin zone boundary\cite{Dai2020,duan2021resonance,raymond2021feedback}, we start from a phenomenological momentum dependence as in Eq. (\ref{eq_chi_phen}). However, now the the momentum dependence contains a one dimensional peak at $\qv_0$, only restricted in $q_x$ and additionally a weaker three dimensional peak at the same momentum,
\begin{align}
 f_{\qv_0}(\qv)=\frac{1}{\frac{\delta q_x^2}{\zeta^2} +1}+\frac{1/2}{\frac{\delta q_x^2+0.1\delta q_y^2+\delta q_z^2}{\zeta^2} +1}
 \label{eq_fq_wall}
\end{align}
where $\delta q_i\equiv q_i-q_{0,i}$.
 \begin{figure}[tb]
  \includegraphics[width=\linewidth]{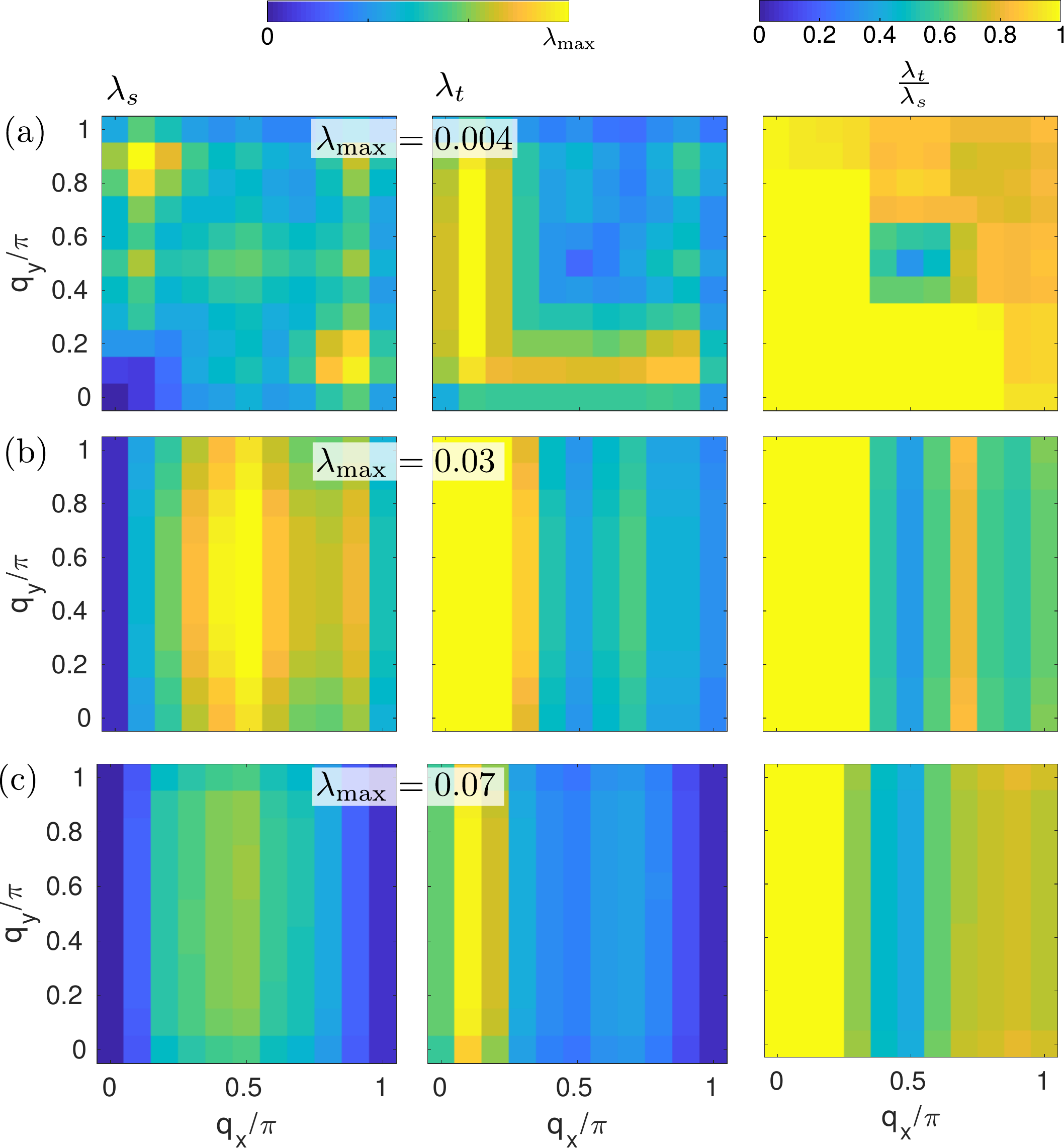}
 \caption{Phase diagram for $q_z=\pi$ using the nontrivial Fermi surfaces for UTe$_2$. (a) Susceptibility peaked at $\qv_0=(q_x,q_y,q_z)$ and symmetry related positions together with the Fermi surface of the 4 band model, $U=0.2$. (b) Susceptibility with elongated peak at $\qv_0=(q_x,q_y,q_z)$ and additional two dimensional contribution at the $q_x$ plane, $U=0.4$, $f_{\qv_0}(\qv)$ in Eq. (\ref{eq_fq_wall}). (c) Susceptibility with peak at $\qv_0=(q_x,q_y,q_z)$ and additional two dimensional contribution at the $q_x$ plane, but using the Fermi surface from the single band model\cite{Shishidou2021}, $U=0.2$. (left: singlet eigenvalue, middle: triplet eigenvalue, right: ratio; maximum of the eigenvalues as indicated).}
 \label{Fig_phase_UTe2}
\end{figure}
 \begin{figure}[tb]
  \includegraphics[width=\linewidth]{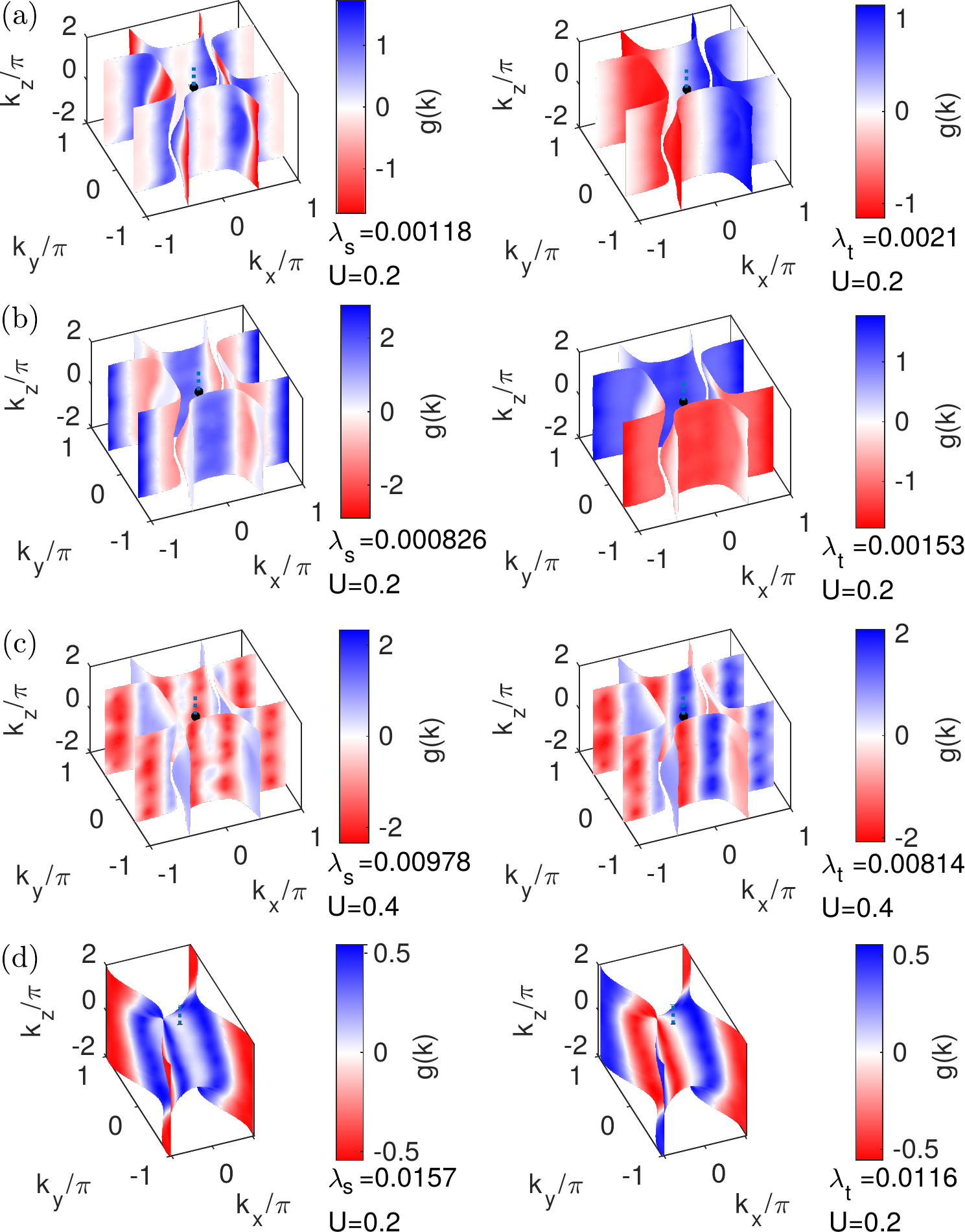}
 \caption{Example pairing states for three dimensional Fermi surface of UTe$_2$. (a) Susceptibility peaked at $\qv_0=(0.1,0.6,1)\pi$ and symmetry related positions. (b) Susceptibility peaked at $\qv_0=(0.6,0.1,1)\pi$ and symmetry related positions. (c) Susceptibility with peak at $\qv_0=(0.7,0,1)\pi$ and additional two dimensional contribution at the $q_x=0.7\pi$ plane, i.e. $f_{\qv_0}(\qv)$ as in Eq. (\ref{eq_fq_wall}). (d) Susceptibility with peak at $\qv_0=(0.7,0,1)\pi$ and additional two dimensional contribution at the $q_x=0.7\pi$ plane, but using the single band Fermi surface.}
 \label{Fig_pairing_UTe2}
\end{figure}
The overall normalization is again chosen such that $U_c=1$ for a magnetic instability at $\qv_0$. Symmetry related peaks for the orthorhombic structure including periodic repetition with respect to the (conventional) Brillouin zone that is chosen to be within $q_x \in [-\pi/a, \pi/a]$, $q_y \in [-\pi/b, \pi/b]$, $q_z \in [-2\pi/c, 2\pi/c]$\footnote{The body centered orthorhombic crystal structure leads to a Brillouin zone in primitive setting where the reciprocal vectors are not pairwise orthogonal. For convenience, we therefore use conventional setting by doubling the volume of the reciprocal space and noting that $\qv^*=(\pi/a,\pi/b,\pi/c)$ connects symmetry related momenta in that volume.} are included. In Fig. \ref{Fig_phase_UTe2} (a), we present a phase diagram as function of the in plane peak position $q_x$ and $q_y$ for a fixed $q_z=\pi$, similar to Fig. \ref{Fig_chi_toy_3d_phase},  using the Fermi surface (including distribution of weights from the Fermi velocity) of our 4 band model (see Fig. \ref{Fig_fermi_UTe2} (a)).
Indeed, because of the orthorhombic symmetry, the phase diagram is less symmetric, and triplet instabilities seem to be favorable only at small $q_x$ and $q_y$. Note that despite the deviations from tetragonal symmetry, the pairing eigenstates are overall very similar for spin fluctuations at $\qv_0=(0.1,0.6,1)\pi$ ($\qv_0=(0.6,0.1,1)\pi$) as demonstrated in Fig. \ref{Fig_pairing_UTe2} (a,b) where the leading triplet states seem to resemble $p_x$ and ($p_y$) pairing instabilities, while the dominant singlet state  exhibits multiple nodes along the $k_y$-$k_z$ ($k_x$-$k_z$) plane.

For a calculation with a susceptibility that resembles the susceptibility obtained from the tight binding models by using the momentum dependence from Eq. (\ref{eq_fq_wall}) with a one dimensional contribution, the phase diagram shows little dependence on $q_y$ as expected from the ridge structure of the susceptibility, see Fig. \ref{Fig_phase_UTe2} (b). The eigenvalues have little dependence on $q_y$, but there appears a region of close competition between singlet and triplet instabilities close to $q_x=0.7\pi$. Pairing states in this region as shown in Fig. \ref{Fig_pairing_UTe2} (c) exhibit accidental nodal lines in the  $k_y$-$k_z$ plane and the triplet solution corresponds to higher harmonics, i.e. the dominant scattering vector connects regions with the same sign of $g(\kv)$ across two nodal planes. Finally, we also calculated a similar phase diagram for the Fermi surface of the single band model
of Ref.\cite{Shishidou2021}, Fig. \ref{Fig_fermi_UTe2} (c) that has overall very similar shape, with a triplet instability that becomes competitive at larger $q_x\approx 0.9\pi$ (Fig. \ref{Fig_phase_UTe2} (c). Indeed the Fermi surface is less two dimensional than for the 4 band model, and the dependence of the eigenvalues $\lambda_s$ and $\lambda_t$ on $q_x$ is less pronounced. Still the eigenstates show nodal lines in the  $k_y$-$k_z$ planes where the positions of the maxima of $|g(\kv)|$ are similar between the singlet and triplet solution; the latter again exhibiting higher order harmonics with the scattering vector across two nodal planes.
Note that the line nodes of the triplet states visible in Fig. \ref{Fig_pairing_UTe2} may disappear in the presence of spin-orbit coupling \cite{Blount1985}.
Let us conclude this discussion of the phenomenological phase diagram by noting that a recent INS investigation\cite{duan2021resonance,raymond2021feedback} has detected dominant spin fluctuations at $(hkl)\approx(0.57,0,0)$ that corresponds to $\qv_0=(1-0.57,0,0.5)2\pi=(0.86\pi,0,\pi)$ in the conventional Brillouin zone setting, i.e. close by to the point in momentum space where competition between singlet and triplet solutions is seen in our phase diagram, see Fig. \ref{Fig_phase_UTe2}. 

\section{Conclusions}

The expectation that UTe$_2$ might be a nearly ferromagnetic superconductor  has not been borne out by inelastic neutron experiments, which show the
magnetic fluctuation spectrum is dominated by a broad  ridge of excitations centered at $\q=2\pi(0.57,0,0)$ in the primitive Brillouin zone.  In most known families of unconventional superconductors, however, finite-$\bf q$ magnetic scattering leads to singlet pairing.  Motivated by this apparent paradox,
we have considered the general question of how 
finite-$\bf q$ spin fluctuations can drive triplet pairing, and further asked if there is anything specific about the UTe$_2$ system that might favor such a scenario.     The spin fluctuation interaction is driven by the magnetic susceptibility $\chi(\q)$, whose peaks are determined by the Fermi surface in single band systems, but may be strongly influenced by higher energy states in the case of multiband superconductors.  We showed that even in one-band situations with very simple, symmetric Fermi surfaces, it is possible that peaks of $\chi(\q)$, calculated in RPA, occur at rather asymmetric positions in $\q$-space, driving triplet superconductivity via an attractive pair scattering connecting order parameters with the same sign.  Such a situation is known to occur, for example, in the one-band Hubbard model quite far from half-filling, which we revisited briefly here.
 
In more complicated  situations, susceptibility peaks can occur at highly asymmetric positions depending on the number of bands, orbital content, and dimensionality of the near-Fermi level electronic structure. We therefore next considered situations with simple Fermi surfaces, but susceptibilities peaked at phenomenologically chosen positions, consistent of course with the symmetries of the underlying crystal.    For such models, we calculated ``phase diagrams" of the amplitude of spin fluctuation pairing in singlet and triplet channels according to the position $q_x,q_y$ of the dominant $\chi(\q)$ peak calculated in the random phase approximation.  In addition to the usual ferromagnetic small-$\bf q$ states, large-$\bf q$ triplet states were found to be competitive or dominant over significant ranges of such phase diagrams.  We discussed what kinds of scattering processes would give rise to such states, and we displayed the structure of the leading gap functions, which generally involve higher-order harmonics of $p$-wave representations, with accidental nodal lines.
 
With specific focus on the UTe$_2$ system, we presented results for a DFT+U calculation of the susceptibility indicating the presence of prominent scattering ridges of intensity aligned along the $y$ direction, qualitatively consistent with INS experiments.  In order to perform the associated spin fluctuation pairing calculations, we used the Fermi surface as obtained from downfolding ab initio results, and again assumed a phenomenological susceptibility inspired by the ab initio calculation.  We found that large-$q$ triplet states are indeed competitive over large parts of the phase diagram, and we compared to results using similar Fermi surfaces obtained from other approaches in the literature.  Thus we conclude that spin fluctuations at finite $q$ may indeed be consistent with the spin triplet pairing deduced phenomenologically by comparison with NMR and other experiments.  
 
We have presented a possible solution to the UTe$_2$ ``paradox" described here within the framework of conventional spin fluctuation theory.  Of course, if the pairing mechanism appropriate to this material turns out to be quite different from what is assumed here, our calculations may not apply to UTe$_2$ directly, but should describe other potential triplet superconductors, including possibly UPt$_3$. In addition, we note that the calculations presented here were for the simplest paramagnetic, spin-rotationally invariant case.  A more complete treatment, including spin-orbit coupling and the associated spin anisotropy in the susceptibility, the orbital degrees of freedom, and a direct calculation of the susceptibility from the model electronic structure, will be postponed to a later date when more certainty exists about the appropriate low-energy model. 
 
\section{Acknowledgements}

P.J.H. was supported by the U.S. Department of Energy under Grant No. DE-FG02-05ER46236. This research was done using resources provided by the OSG \cite{osg07,osg09}, which is supported by the National Science Foundation award \#{}2030508.

\bibliography{bibliography_triplet}
\ifarXiv
    

\section*{Supplementary material}

# Supplementary Material: Spin triplet superconductivity driven by finite momentum spin fluctuations

Andreas Kreisel,[1] Yundi Quan,[2] and P. J. Hirschfeld[2]

[1]*Institut für Theoretische Physik Universität Leipzig D-04103 Leipzig, Germany*
[2]*Department of Physics, University of Florida, Gainesville, Florida 32611, USA*

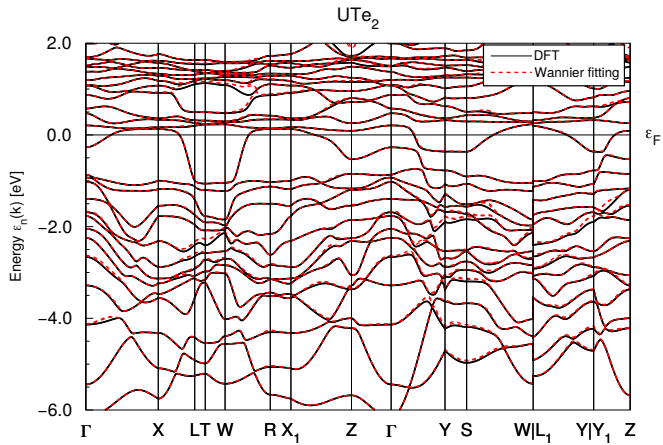

FIG. S1. The tight-binding band structure (red) can reasonably reproduce the DFT band structure near the Fermi level.

## I. COMPUTATIONAL DETAILS AND CRYSTAL STRUCTURE OF UTe$_2$

First-principles calculations are carried out using the full-potential local-orbital code FPLO [1, 2]. We used the local density approximation (LDA) as the exchange correlation potential [3] and the Kohn-Sham equation is solved in fully relativistic basis. The space group of UTe$_2$ is *Immm* (#71) and the lattice constants $a$, $b$ and $c$ are 4.161, 6.122, 13.955 Å respectively with the atomic positions as given in Table S1.[4, 5] We have used the

LDA+U method with an effective $U_{eff}$=2 eV and $J = 0$, which corresponds to $F0 = 2eV$ and $F2 = F4 = F6 = 0$ yielding a good agreement between the calculated Fermi surfaces and those measured in ARPES[6]; a similar calculation was already done in Ref. [7]. We have used a $12 \times 12 \times 12$ mesh for self-consistent calculations and imposed the charge convergence criteria of $10^{-6}$.

Uranium 5f $J = \frac{7}{2}$ states lie about 1 eV above the Fermi level and are relatively dispersiveless, see Fig. S3 In contrast, the $J = \frac{5}{2}$ states are hybrized with Te2 5p states near the Fermi level, see Fig. S2 and Fig. S4.

| | $x$ | $y$ | $z$ | Wyckoff site |
|---|---|---|---|---|
| U | 0 | 0 | 0.1354 | 4i |
| Te1 | 0.5 | 0 | 0.2975 | 4j |
| Te2 | 0.5 | 0.2491 | 0.0 | 4h |

TABLE S1. Atomic positions in terms of internal coordinates. Note that there are two Te sites.

## II. 56-BAND TIGHT-BINDING MODEL OF UTe$_2$

To accurately reproduce the electronic structure of UTe$_2$ near the Fermi level, we construct a tight-binding model with 56 orbitals that consists of U d, f and Te p orbitals. In Fig. S1, we compare the DFT band structure with the tight-binding band structure and they agree reasonably well near the Fermi level. The fatbands for the U 5f $|J = 5/2\rangle$ are shown in Fig. S2 and the ones for U 5f $|J = 7/2\rangle$ in Fig. S3 respectively.

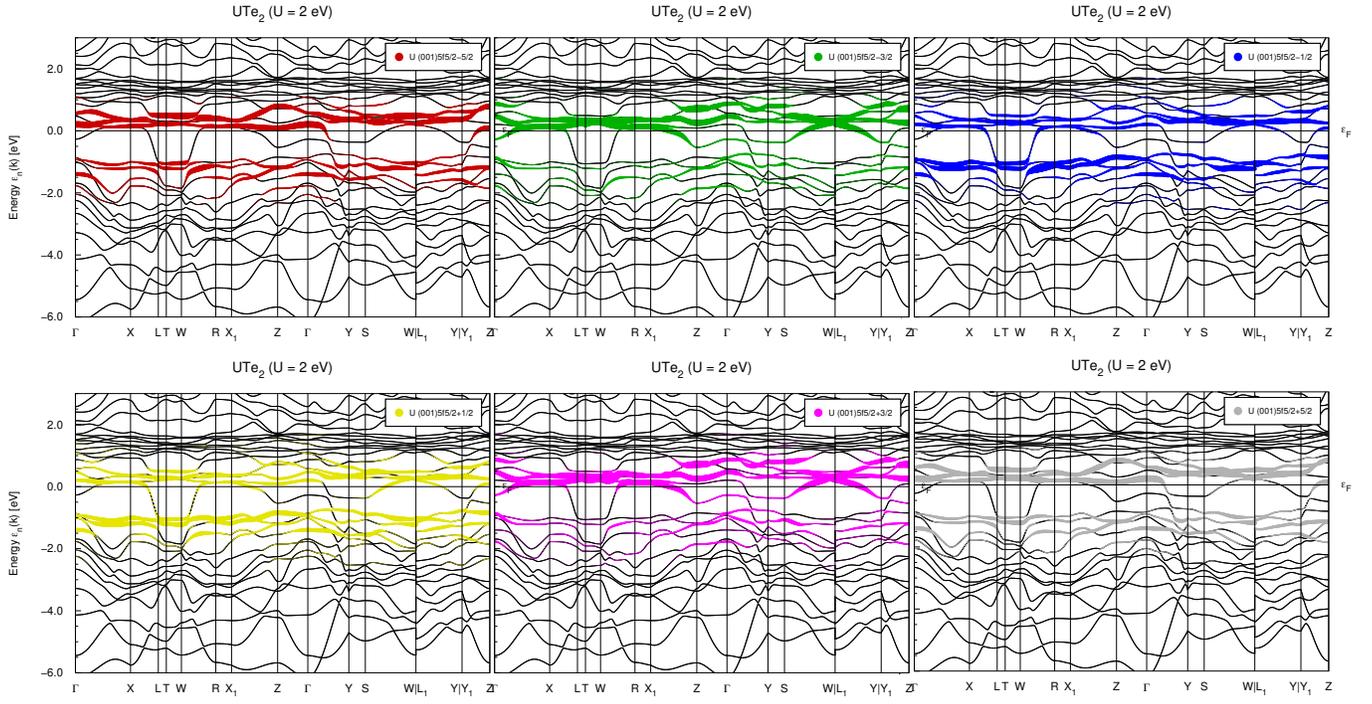

FIG. S2. Fatbands showing the partial weight for the individual U 5f $|J = 5/2\rangle$ states, see caption.

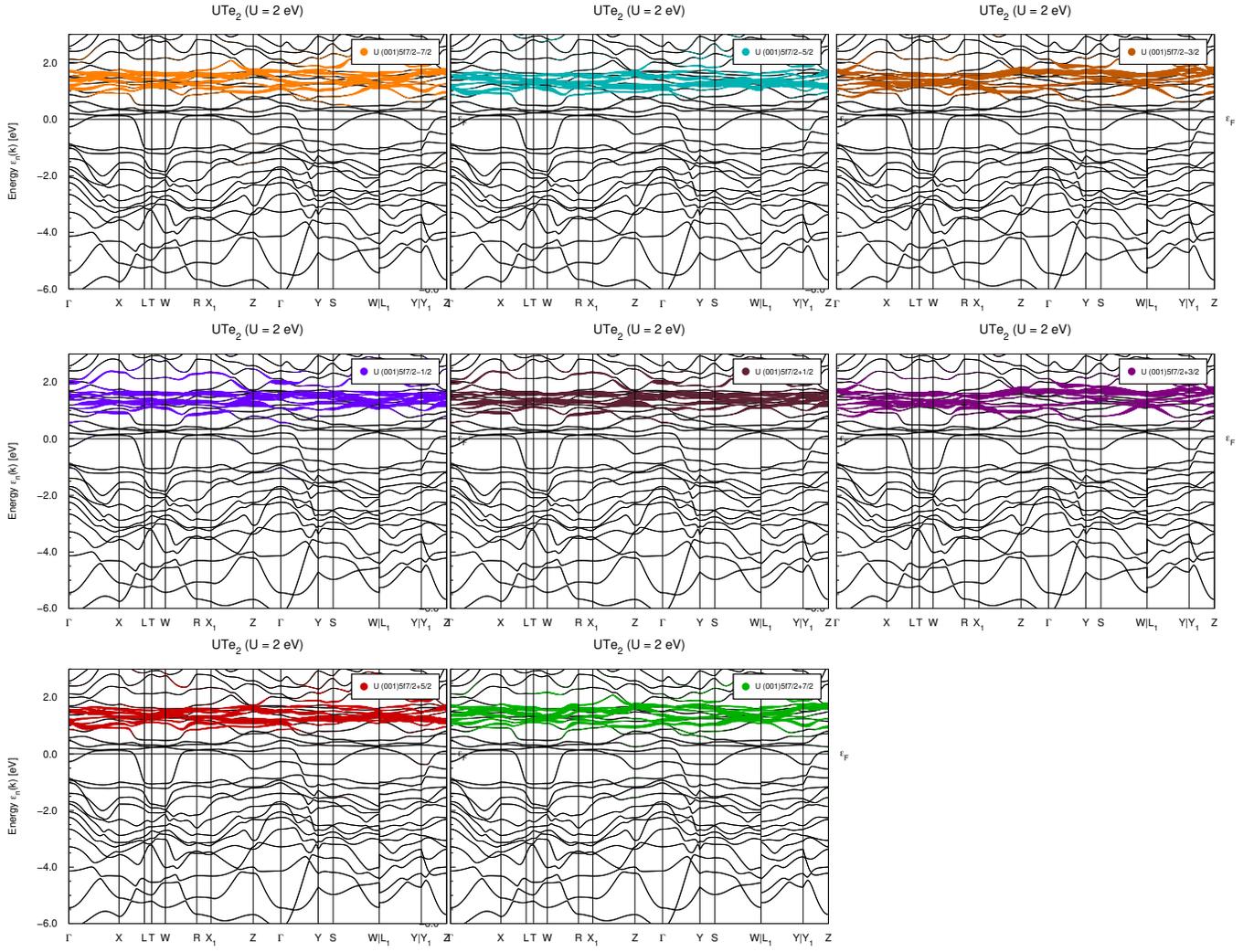

FIG. S3. Fatbands for U 5f $|J = 7/2\rangle$ states.

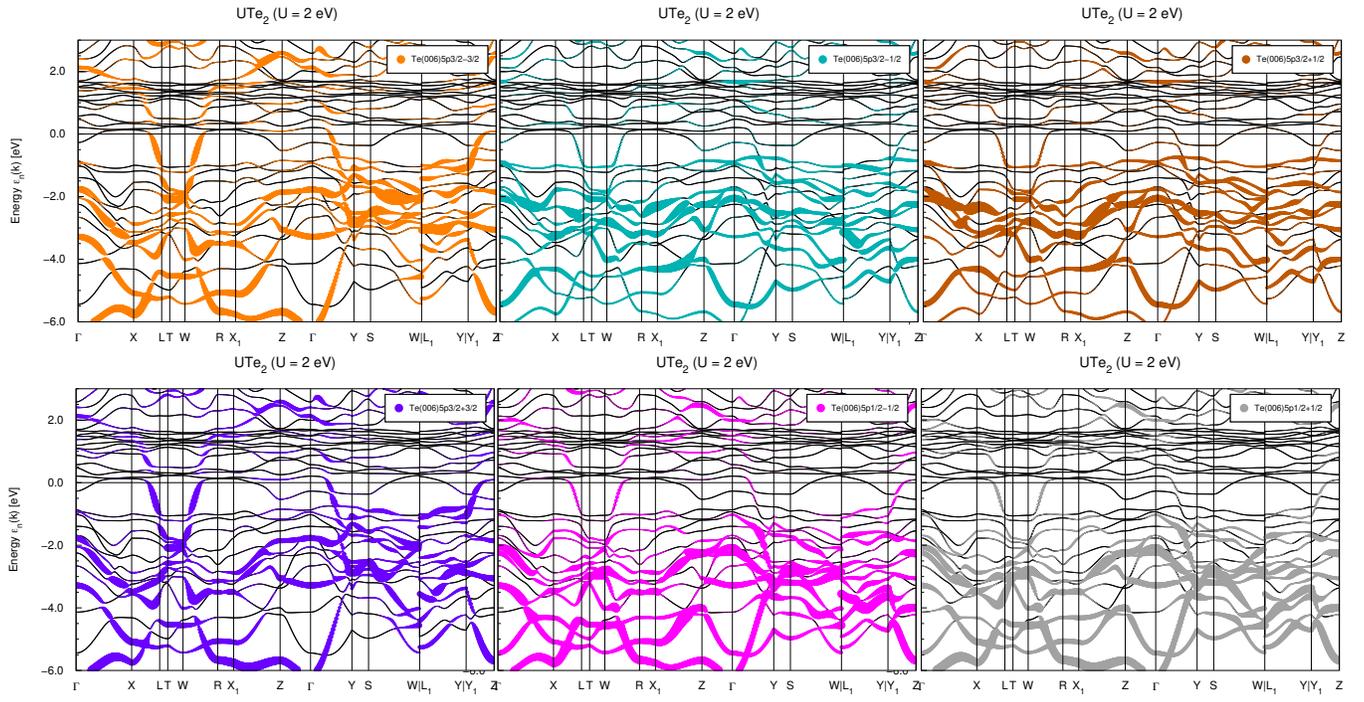

FIG. S4. Fatbands of Te2 5$p$ orbitals.


\fi
\end{document}